\documentclass[manuscript]{aastex}
\usepackage{natbib}
\usepackage[stable,symbol]{footmisc}
\usepackage{hyperref}

\shorttitle{SED of the coolest Y dwarf}
\shortauthors{Kopytova et al.}

\begin{document}


\title{Deep $z$-band observations of the coolest Y dwarf\thanks{Based on observations made with ESO Telescopes at the La Silla Paranal Observatory under program 293.C-5008(A)}}


\author{Taisiya G. Kopytova\altaffilmark{1,2,3,+}, Ian J. M. Crossfield\altaffilmark{1,7,\times}, Niall R. Deacon\altaffilmark{1}, Wolfgang Brandner\altaffilmark{1}, Esther Buenzli\altaffilmark{1}, Amelia Bayo\altaffilmark{1,4},  Joshua E. Schlieder\altaffilmark{1,6}, Elena Manjavacas\altaffilmark{1,2}, Beth A. Biller\altaffilmark{5}, and Derek Kopon\altaffilmark{1}}

\email{kopytova@mpia.de}

\affil{\altaffilmark{1}Max Planck Institute for Astronomy, K\"{o}nigstuhl 17, 69117 Heidelberg, Germany}

\affil{\altaffilmark{2}International Max Planck Research School for Astronomy and Space Physics, Heidelberg, Germany}

\affil{\altaffilmark{3}Astrophysics Department, American Museum of Natural History, 200 Central Park West, New York, NY 10024, USA }

\affil{\altaffilmark{4}Instituto de Fisica y Astronom\'{i}a, Facultad de Ciencias, Universidad de Valpara\'{i}so, Gran Breta\~{n}a N¼ 1111, Playa Ancha, Valpara\'{i}so, Chile}

\affil{\altaffilmark{5}Institute for Astronomy, University of Edinburgh, Blackford Hill View, Edinburgh EH9 3HJ, UK}

\affil{\altaffilmark{6}NASA Postdoctoral Program Fellow, NASA Ames Research Center, Space Science and Astrobiology Division, MS 245-6, Moffett Field, CA 94035, USA}

\affil{\altaffilmark{7}Lunar and Planetary Lab, 1629 E University Blvd, Tucson, AZ 85721, USA}

\altaffiltext{+}{Visiting Scholar, Kade Fellow at the American Museum of Natural History}
\altaffiltext{$\times$}{Sagan Fellow}




\begin{abstract}
WISE J085510.83-071442.5 (hereafter, WISE 0855-07) is the coolest Y dwarf known to date and is located at a distance of 2.31$\pm 0.08$ pc, giving it the fourth largest parallax of any known star or brown dwarf system. We report deep $z$-band observations of WISE 0855-07 using FORS2 on UT1/VLT. We do not detect any counterpart to WISE 0855-07 in our $z$-band images and estimate a brightness upper limit of AB mag $>$ 24.8 ($F_{\nu}$ $<$ 0.45 $\mu$Jy) at 910 $\pm$ 65 nm with $3\sigma$-confidence. We combine our z-band upper limit with previous near- and mid-infrared photometry to place constraints on the atmospheric properties of WISE 0855-07 via comparison to models which implement water clouds in the atmospheres of $T_{eff} < 300$ K substellar objects. We find that none of the available models that implement water clouds can completely reproduce the observed SED of WISE 0855-07. Every model significantly disagrees with the (3.6 $\mu$m / 4.5 $\mu$m) flux ratio and at least one other bandpass. Since methane is predicted to be the dominant absorber at 3--4 $\mu$m, these mismatches might point to an incorrect or incomplete treatment of methane in current models. We conclude that \mbox{(a) WISE0855-07} has $T_{eff} \sim 200-250$~K, (b) $< 80 \%$ of its surface is covered by clouds, and (c) deeper observations, and improved models of substellar evolution, atmospheres, clouds, and opacities will be necessary to better characterize this object.
\end{abstract}





\section{Introduction}

Y dwarfs are substellar objects located at the coolest and lowest-mass edge of the brown dwarf M-L-T-Y spectral sequence \citep[][and ref. therein]{Kirkpatrick_etal_2012}.
Previous studies of Y0-1 spectral type objects reveal effective temperatures of \mbox{400--500 K} and masses of 10--30 M$_{Jup}$ \citep[e.g.][]{Cushing_etal_2011, Dupuy_Kraus_2013, Leggett_etal_2013}.
Models predict  a distinct atmospheric chemistry for Y dwarfs: NH$_{3}$ becomes apparent in the near-infrared and various species condense to form clouds \citep{Burrows_etal_2003}.
In the warmest Y dwarfs these clouds are composed of Na$_{2}$S \citep{Morley_etal_2012}, but at temperatures lower than 300 K the clouds may include H$_{2}$O, NH$_{3}$, 
and other, more exotic species \citep{Burrows_etal_2003, Visscher_etal_2006, Morley_etal_2014}.

Most recently, \cite{Luhman_2014} announced the detection of WISE J085510.83-071442.5 (hereafter, WISE 0855-07), a Y dwarf with $T_{eff}=235-260 $ K at a distance of 2.31$\pm 0.08$ pc \citep{Luhman_Esplin_2014}, making it the fourth closest known stellar or brown dwarf system. WISE 0855-07 is the coolest Y dwarf known to date. Occupying a temperature regime intermediate between hotter and more massive L and T dwarfs ($T_{eff}$ $ >550 $ K) and Jupiter \citep[$T_{eff}=126$ K; e.g.][]{Li_etal_2012}, WISE 0855-07 provides a unique opportunity to test the presence of water clouds in the atmospheres of Y dwarfs at temperatures \mbox{below 400 K}. 

\citet{Luhman_2014} and \citet{Wright_etal_2014} report observations of WISE 0855-07 from WISE \citep{Wright_etal_2010}, Spitzer, and ground-based facilities.  However, the object is only detected by WISE and Spitzer in essentially two bandpasses at 3.6 $\mu$m and 4.5 $\mu$m.  \citet{Beamin_etal_2014} also report a non-detection at $Y$ band giving an upper limit of \mbox{$Y>24.4$ mag} at the 3$\sigma$-level. On the other hand, \citet{Faherty_etal_2014} announce a 2.6$\sigma$-detection of WISE 0855-07 giving $J3 = 24.8^{+0.53}_{-0.35} (J_{MKO} = 25.0^{+0.53}_{-0.35}$), or equivalently an upper limit of $J3 > 23.8$ mag ($J_{MKO} > 24.0$ mag) at 5$\sigma$. \citeauthor{Faherty_etal_2014} compare these observations with chemical equilibrium atmospheric models and demonstrate that WISE 0855-07 is 2.7$\sigma$ from cloudless atmosphere models and can be reproduced by partly cloudy models (50\%) containing sulfide and water ice clouds. However, the latter have been disputed by \citet{Luhman_Esplin_2014} who find that the SED of WISE 0855-07 can be explained by cloudless models that implement non-equlibrium chemistry. Nevertheless, we show that none of the available models that implement water clouds match all the existing observations completely (see Section 3), emphasizing that theoretical predictions for Y dwarf atmospheres are still quite uncertain. Obtaining a complete SED for WISE 0855-07 is essential for understanding atmospheric properties in the temperature regime below 400 K. In this paper, we present deep $z$-band observations of WISE 0855-07, determine its upper-brightness limit from our non-detection and discuss how our result constrains this object's atmospheric properties.

\section{Observations and data analysis}
Observations of WISE 0855-07 were carried out on May 31, 2014 using FORS2 mounted on VLT/UT1 at the ESO/Paranal observatory in Chile. FORS2 is a visual and near-UV focal reducer and low-dispersion spectrograph \citep{Appenzeller_etal_1998}. Observation were obtained in imaging mode using the red-optimized CCD through the z$\_$GUNN+78 filter ($\lambda_0$=910 nm, FWHM=130.5 nm) with the high resolution collimator that gives a field of view of $4\farcm2 \times 4\farcm2$. Pixels were binned (2$\times$2) resulting in a pixel scale of $0\farcs125$/pixel. In total, six images with an exposure time of 480 s each were taken. The telescope was pointed so that the predicted position of WISE 0855-07 was located on the upper chip of FORS2 which has a better sensitivity. A small telescope offset was applied after each integration, to avoid bad pixels and cosmic ray contamination.

All six frames were reduced in the standard manner using IRAF\footnote{IRAF is distributed by the National Optical Astronomy Observatories,
    which are operated by the Association of Universities for Research
    in Astronomy, Inc., under cooperative agreement with the National
    Science Foundation.} \citep{Tody_1993} - bias subtraction, flat-fielding, fringe correction and sky subtraction. The reduced frames were aligned and co-added, in order to obtain a higher signal-to-noise. Astrometry and photometry from the Pan-STARRS1 (PS1) catalog \citep{Schlafly_etal_2012, Torny_etal_2012, Magnier_etal_2013} were used to measure the world coordinate system and to provide flux calibrations for the combined image. We used the method of \mbox{Finkbeiner et al. 2014 (submitted to ApJ)} to transform $z_{PS1}$ to the standard $z_{SDSS}$. Due to the lack of standard stars with z$\_$GUNN+78 measurements, we cannot apply a proper transformation between z$\_$GUNN+78 and $z_{SDSS}$ magnitudes. However, we apply the filter responses of the $z$\_GUNN+78\footnote{Available through the FORS2 Exposure Time Calculator at \url{http://www.eso.org/observing/etc/}} and $z_{SDSS}$ filters to theoretical spectra \citep[][see Section 3]{Burrows_etal_2003, Morley_etal_2014} and find that the resulting flux varies at most $\pm0.05$ mag from filter to filter. Hence, the uncertainty in the magnitude system cannot account for the differences between the observed data and the models (see Section 3).

\begin{figure}
\centering
\includegraphics[scale=0.5]{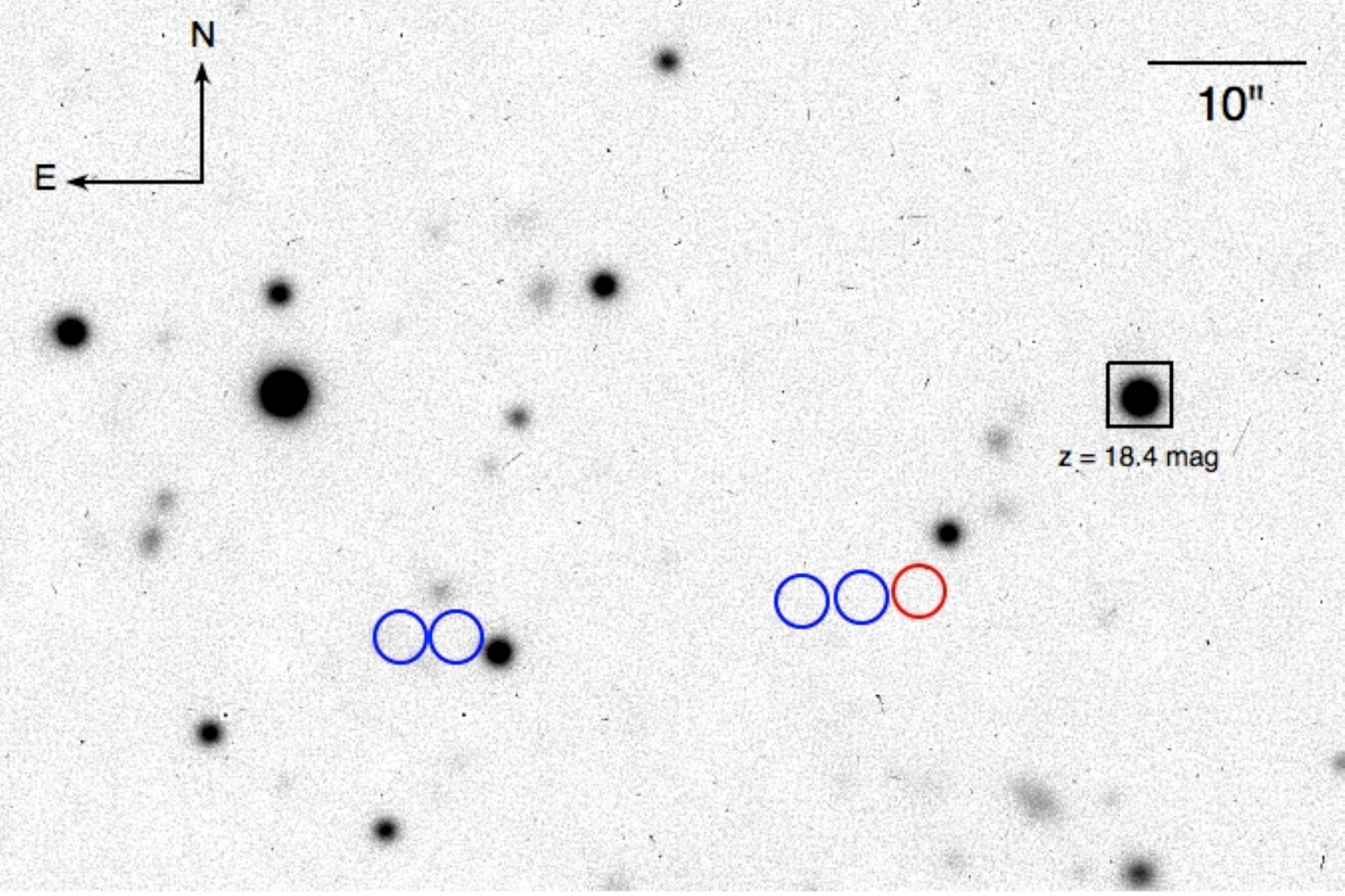}
\caption{$z$-band FORS2 observations of the field containing WISE0855-07. Blue circles are WISE positions in 2010.34 and 2010.86 (the pair on the left) and Spitzer positions 2013.47 and 2014.05 (the pair on the right). The red circle is the expected position of WISE 0855-07 on May 31, 2014. No counterpart to WISE 0855-07 is detected in our $z$-band FORS2 frames. For reference, the square indicates a source with $z$(AB)=18.4 mag. \label{Fig_1} }
\end{figure}

We used IRAF/DAOFIND to search for a counterpart to WISE 0855-07. DAOFIND approximates a stellar point spread function with an elliptical Gaussian function. DAOFIND identifies no counterpart to WISE 0855-07 at its expected position in our z-band frames (Fig. \ref{Fig_1}). To place an upper limit on the z-band magnitude of the brown dwarf, we estimate the sky brightness and the sky standard deviation at this expected position. Using the calibrated photometry from PS1, we estimate an upper brightness limit for WISE 0855-07 of $z_{AB} > 24.8$ mag, or $F_{\nu}<0.45 \mu$Jy with 3$\sigma$-confidence (corresponding to a direct measurement and uncertainty of $0.06\pm0.13 \mu$Jy). 
This result is consistent with the non-detection of WISE 0855-07 in the $Y$-band using HAWK-I on UT4/VLT at the Paranal observatory  by \citep{Beamin_etal_2014} and 2.6$\sigma$-detection in $J$ band by \citep{Faherty_etal_2014}.





\section{Comparison with models and discussion}

\begin{figure}
\centering
\includegraphics[scale=0.7]{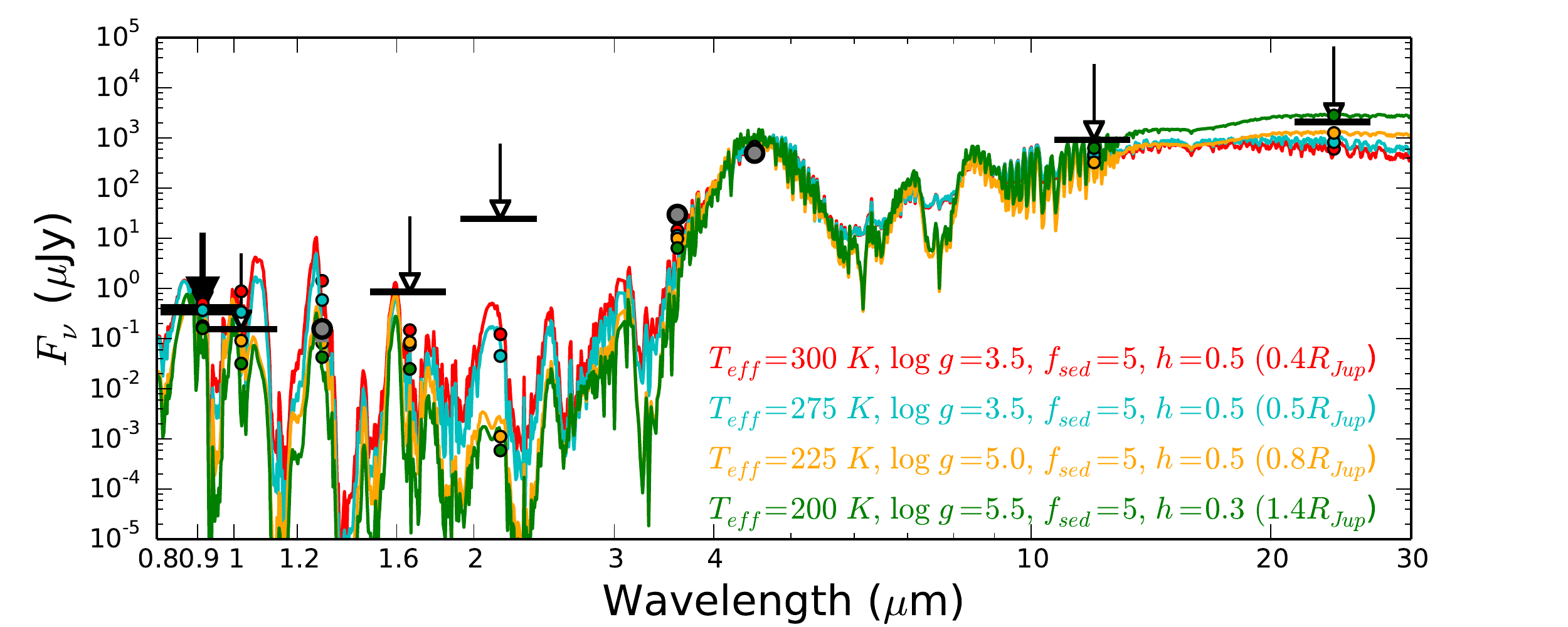}
\caption{\citet{Morley_etal_2014} atmospheric models for various effective temperatures $T_{eff}$, surface gravities $\log g$, and cloud sedimentation parameters $f_{sed}$. Large
grey filled circles are detections using WISE \citep{Wright_etal_2010} and Spitzer, and the 2.6$\sigma$ $J$-band detection by \citet{Faherty_etal_2014}. Large arrows are upper brightness limits from \citet{Luhman_2014}, \citet{Wright_etal_2014} and \citet{Beamin_etal_2014}. The small color-coded dots are fluxes in photometric bands predicted by the models. Our upper limit is highlighted at far left. All photometric measurements are on the AB system. The hottest model shown is ruled out both by the upper limits and by the physically implausible radius required to match the detections.
\label{morley}}
\end{figure}

\begin{figure}
\centering
\includegraphics[scale=0.7]{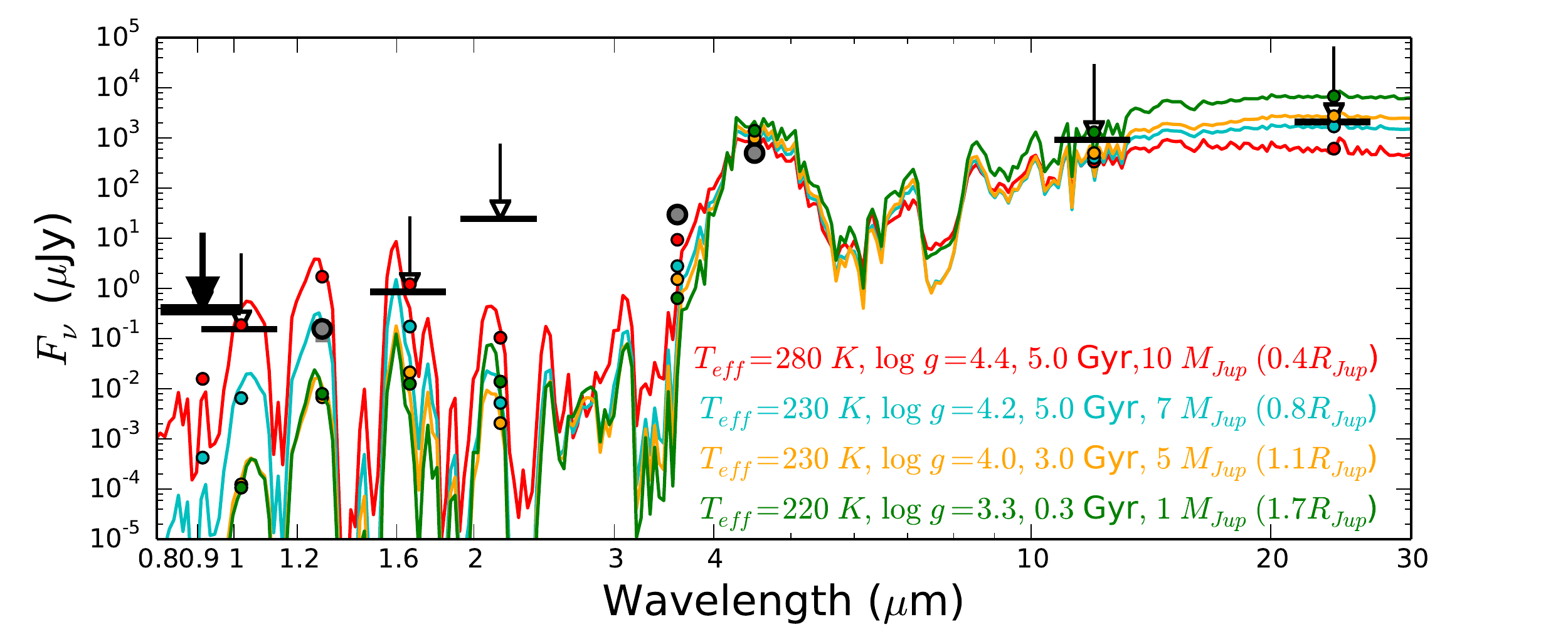}
\caption{\citet{Burrows_etal_2003} atmospheric models for various effective temperatures $T_{eff}$, surface gravities $\log g$ and ages. Large
grey filled circles are detections using WISE \citep{Wright_etal_2010} and Spitzer, and  the 2.6$\sigma$ $J$-band detection by \citet{Faherty_etal_2014}. Large arrows are upper brightness limits from \citet{Luhman_2014}, \citet{Wright_etal_2014} and \citet{Beamin_etal_2014}. The small color-coded dots are fluxes in photometric bands predicted by the models. Our upper limit is highlighted at far left. All photometric measurements are on the AB system. The hottest model shown is ruled out both by the upper limits and by the physically implausible radius required to match the detections.\label{burrows} }
\end{figure}

We compare the ensemble of observations of WISE 0855-07 to atmospheric spectral models of cool substellar objects.  We considered the full set of model spectra from \citet{Morley_etal_2014} and \citet{Burrows_etal_2003}, both of which extend well below 300 K and include the effects of water clouds. Figures \ref{morley} and \ref{burrows}  show the observations and several selected models with $T_{eff}$ ranging from 200 to 300~K.   For each model, we compute the flux expected in the Spitzer/IRAC 3.6 $\mu$m and 4.5 $\mu$m channels and scale the models to the observed Spitzer fluxes using the approach described by \citet{Rayner_etal_2009}.  We convert these scale factors to physical radii, which are listed in the figure legend.

None of the models match all the data, but models with $T_{eff}\lesssim 250$~K give the most reasonable agreement. Most notable is that no model faithfully reproduces the observed [3.6] - [4.5] color; models hotter than 300~K begin to match this color, but predict optical/NIR fluxes that would have been easily detected. Moreover, the physical radii of 0.5 and 0.6 R$_{Jup}$ required to fit 300 K \citeauthor{Morley_etal_2014} and 280 K \citeauthor{Burrows_etal_2003} models, respectively, are smaller than radii predicted from equations of state of very low-mass objects. A larger coverage fraction of cold clouds for models of $T_{eff} = 300$ K would give more reasonable radii, but disagrees with the upper limits. The \citeauthor{Burrows_etal_2003} models show a significant discrepancy with the \citeauthor{Morley_etal_2014} models at $\lambda <1.2 \, \mu$m. The best-fitting \citeauthor{Morley_etal_2014} 250~K models and cooler, heavily-clouded \citep[$h\gtrsim0.8$ in the nomenclature of][]{Marley_etal_2010, Morley_etal_2014} models predict $z$-band fluxes above our detection limit, so our observations nominally exclude these models.   Although atmospheric models with $T_{eff}=200$~K and a partly-cloudy atmosphere ($h=0.5$) agree with most published upper limits, they formally disagree with the upper limit of W4 $<$ 9 mag reported by \citet{Luhman_2014}.   Models below $\approx$200~K have [4.5] -- [W4] colors that are excluded by existing data. The coolest models plotted in Figures 2 and 3 predict radii of 1.4 $R_{Jup}$; this value is larger than predicted by evolutionary models, but would be consistent with an unresolved, near-equal-mass binary. A similar situation holds with the models of \citet{Burrows_etal_2003}, shown in Fig.~3, where the hotter models are too bright in the NIR and the cooler models are too bright at W4.  Thus, every model disagrees with the 3.6 $\mu$m point and at least one other bandpass. The methane molecular band is the most significant opacity source at  3.6--4.5  $\mu$m, therefore the inconsistency between theoretical predictions and observations at this wavelength range might be explained by missing and/or incorrect line opacities. Alternatively, the mismatch between observations and models could indicate a more shallow temperature-pressure profile than predicted by current theory or by non-equilibrium chemistry (e.g. vertical mixing) as suggested for some directly imaged giant exoplanets \citep[e.g.][]{Skemer_etal_2014}.

Based on the evolutionary models of \citet{Saumon_Marley_2008}, at an age of 10~Gyr a 200~K object must have a mass $\le$ 15 M$_{Jup}$ and radius $\approx$ 1.0 R$_{Jup}$.  These constraints suggest that WISE 0855-07 should have $\log g \lesssim 4.5$; the plotted model spectrum with the faintest optical/NIR fluxes corresponds to $\log g = 5.0$, which would be formally excluded based on evolutionary considerations.  Nonetheless even this model predicts a $z$-band flux only $\sim$4$\times$ fainter than our limit, and the model with 200~K and $\log g$=4 yields a $z$-band flux barely fainter than our new constraint; thus dedicated ground-based observations still have a role to play.  Our primary conclusions are therefore that (a) based on the current state-of-the-art models, WISE0855-07 has $T_{eff} \lesssim 250$~K and (if patchy) $\lesssim$80\% of the surface is cloud-covered, and (b) improvements in models of substellar evolution, atmospheres, clouds, and opacities will be necessary to better characterize this object. WISE 0855-07 has the potential to become the first object outside the Solar system with detected water clouds in its atmosphere.

\acknowledgments

\textit{Acknowledgements.}
We gratefully acknowledge the ESO Director's Discretionary Time Committee for awarding DDT time to this project.
We thank the Paranal observatory staff for conducting the service mode observations presented here. This research has made use of the SIMBAD database, operated at CDS, Strasbourg, France.

We thank the Pan-STARRS1 Builders and Pan-STARRS1 operations staff for construction and operation of the Pan-STARRS1 system and access to the data products provided.The Pan-STARRS1 Surveys (PS1) have been made
possible through contributions of the Institute for Astronomy, the University of Hawaii, the Pan-STARRS
Project Office, the Max-Planck Society and its participating institutes, the Max Planck Institute for Astronomy, Heidelberg and the Max Planck Institute for Extraterrestrial Physics, Garching, the Johns Hopkins University, Durham University, the University of Edinburgh,
QueenÕs University Belfast, the Harvard-Smithsonian
Center for Astrophysics, the Las Cumbres Observatory
Global Telescope Network Incorporated, the National
Central University of Taiwan, the Space Telescope Science Institute, the National Aeronautics and Space Administration under Grant No. NNX08AR22G issued
through the Planetary Science Division of the NASA Science Mission Directorate, the National Science Foundation under Grant No. AST-1238877, the University of
Maryland, and Eotvos Lorand University (ELTE).

TK is partly supported by a Kade Fellowship at the American Museum of Natural History (AMNH). EB is supported by the Swiss National Science Foundation (SNSF). A portion of the research of J.E.S. was supported by an appointment to the NASA Postdoctoral Program at NASA Ames Research Center, administered by Oak Ridge Associated Universities through a contract with NASA.
TK is grateful to Juan Carlos Beamin and Jackie Faherty for helpful discussions regarding WISE 0855-07.



{\it Facilities:} \facility{VLT:Antu, Pan-STARRS1}.



\clearpage

\appendix

\clearpage

\clearpage






\end{document}